# High performance of mixed halide perovskite solar cells: role of halogen atom and plasmonic nanoparticles on the ideal current density of cell

Mohammad Ali Mohebpour, Mohaddeseh Saffari, Hamid Rahimpour Soleimani, Meysam Bagheri Tagani[*]

**Computational Nanophysics Laboratory (CNL), Department of Physics, University of Guilan, PO Box 41335-1914, Rasht, Iran**

*Corresponding Author, Email: m_bagheri@guilan.ac.ir

**Abstract**

To be able to increase the efficiency of perovskite solar cells which is one of the most substantial challenges ahead in photovoltaic industry, the structural and optical properties of perovskite $CH_3NH_3PbI_{3-x}Br_x$ for values x=1-3 have been studied employing density functional theory (DFT). Using the optical constants extracted from DFT calculations, the amount of light reflectance and ideal current density of a simulated single-junction perovskite solar cell have been investigated. The results of DFT calculations indicate that adding halogen bromide to $CH_3NH_3PbI_3$ compound causes the relocation of energy bands in band structure which its consequence is increasing the bandgap. In addition, the effect of increasing Br in this structure can be seen as a reduction in lattice constant, refractive index, extinction and absorption coefficient. As well, results of the simulation suggest a significant current density enhancement as much as 22% can be achieved by an optimized array of Platinum nanoparticles that is remarkable. This plan is able to be a prelude for accomplishment of solar cells with higher energy conversion efficiency.

**Introduction**

The efficiency enhancement of perovskite solar cells in short time frame in recent years, have been considered by researchers and scientists of photovoltaics industry and it seems to be a wonderful and practical matter for studying. The enhancement efficiency from about 3.8% in 2009 [1] to 22.1% in 2016 [2-4] for this matter has reported which among the other generations of solar cells is unique and demonstrates the high potential for optoelectronic applications [5,6]. To exploit the solar cells more and better, first it is necessary that we have comprehensive information on the microscopic properties of absorbent layer. The results of experimental measurements show that due to temperature, pressure or external field changes [7,8] and variation of type and size in organic cation [9], phase transition [8,10] and consequently the structural properties would change [11]. The perovskite is in cubic phase in high temperatures [12], in orthorhombic one in low temperatures [13] and in tetragonal phase in moderate temperature [14].

Berdiyorov et al studied the effect of perovskite on electrical and structural properties of compounds and found that structures with more symmetry such as cubic phase were more desirable [11]. The bandgap of perovskite is not the same in different phases [11,15]. General structure of perovskite is ABX3 that A cation has basic and determinant role on the properties of perovskites and efficiency of perovskite solar cells. Also, it can be monovalent metal or an organic molecule. One of the most efficient solar cells that is related to methylammonium lead iodine ($CH_3NH_3PbI_3$) in which A is organic molecule ($CH_3NH_3$). Interestingly, the organic cation can rotate in structure and take different directions [16,17]. Rotation needs energy [7] so that it has a specific direction in orthorhombic phase [8,10]. But in high temperature and cubic phase due to increase of molecule energy, there is no barrier against movement and rotation of molecule and it is so fast that one cannot consider specific direction for it [10]. Movement and rotation of organic cation in perovskite compound is effective on the structural properties as bandgap of these structures [11,17] and causes the bandgap to convert from direct to indirect [17]. For this reason, we call them dynamical band gap semiconductors [17]. Changing direction of organic molecule causes direct bandgap at R point and shifts to Γ point as an indirect bandgap that shows valance band minimum and conduction band maximum dependent to rotation and space alignment of molecule. Long lifetime of charge carriers is also attributed to the dynamical band gap of the perovskites. This causes the molecules to move and prevent the carrier recombination that directly has a positive effect on the enhancement of efficiency [17].

The unique features of perovskites that has become the source of focus for researchers at the moment are as follows: suitable bandgap about 1.5 eV [15] and significant light absorption, separation of carriers in perovskite and ability of transferring charge carriers in the absence of hole transport [18,19], long excitonic diffusion length [20] and high mobility of charge carriers [21-23]. Another feature of this matter is suitable thermoelectric properties including ultra-low thermal conductivity [24]. Perovskite is employed for making of electronic devices such as diodes [25]. For efficient use of sunlight, we should make the solar cell with high efficiency, low cost and high stability. Although perovskite solar cells show high efficiency, high flexibility [26] and low manufacturing cost [27], but they do not have desirable stability and there is an instability problem for them. One of the ways for improvement the stability of matter is doping with other atoms. Adding Sn to Pb in $CH_3NH_3PbI_3$ compound or doping with Br and Cl halogens [15,28-34].

Jong et al. in their work showed that by adding Br to $CH_3NH_3PbI_3$ from 0 to 1, $CH_3NH_3Pb(I_{1-x}Br_x)_3$, the best state with proper efficiency and stability could be obtained for Br=0.2 [34]. In the investigations done by Berdiyorov et al. in tetragonal phase, more transmission and better I-V curve obtained for doped compound with one bromide halogen ($CH_3NH_3PbI_2Br$) [29]. Also one should be careful in selection of amount of doping since Berdiyorov et al. showed that full halide $CH_3NH_3PbCl_3$ and $CH_3NH_3PbBr_3$ causes decrease of electronic transport and current in system [29].

To create a typical perovskite solar panels, at least five different layers will be required including anti-reflection coating, transparent conductive oxides (TCO), hole and electron transport layer (HTL and ETL) and metal back contact with high electrical conductivity [35]. In such a structure, the active layer is sandwiched between two transport layers (HTL and ETL). The main way to increase the efficiency of these solar cell and obtain the maximum output is to improve the optical absorption. To accomplish this task, there are several influential mechanisms in which using plasmonic nanoparticles and light trapping design are the most efficient [36,37]. In majority of experimental and theoretical papers, Ag and Au are known as a cost-efficient material with considerable plasmonic effects [38,39] while Platinum nanoparticles exhibit localized surface plasmons with a higher sensitivity as well [40-42].

In this research, we check out the structural and optical properties of perovskite $CH_3NH_3PbI_{3-x}Br_x$ for values x=1-3 in cubic phase by employing density functional theory. Later, we optically model a finite-sized perovskite solar cell with different plasmonic arrays based on a full solution to Maxwell's equations, using FDTD techniques. After optimization the whole cells parameters, a significant ideal current enhancement as much as 22% is found which is beyond the existing record for perovskite cells.

**Method of Calculations**

DFT calculations: analysis of structural and optical properties of considered perovskites performed by using density function theory. Initially, the unit cells were optimized via Quantum Espresso software [43] using k point sampling of (8× 8× 8). OB86-VWD-DF method [44] was used to consider wan der walls interactions which are essential in perovskite structures. Siesta package [45] was utilized to compute density of states by k-point sampling of (8× 8× 8). GGA-PBE approximation and double-zeta-double-polarized basis sets were employed to describe the valance electrons. The mesh cutoff energy in computation is 200Ry. Calculations related to optical properties have been done by Siesta package by k-point mesh of (50× 50× 50). Optical constants such as refractive index and extinction coefficient have been determined by using real and imaginary parts of dielectric constant ($\varepsilon = \varepsilon_1 + i\varepsilon_2$) as the following:

$$n(\omega) = \frac{1}{\sqrt{2}} \left( \sqrt{\varepsilon_1^2(\omega) + \sqrt{\varepsilon_1^2(\omega) + \varepsilon_2^2(\omega)}} \right) \quad (1)$$

$$k(\omega) = \frac{1}{\sqrt{2}} \left( \sqrt{\varepsilon_1^2(\omega) - \sqrt{\varepsilon_1^2(\omega) + \varepsilon_2^2(\omega)}} \right) \quad (2)$$

FDTD simulation procedure for perovskite solar cell: In this work, a single-junction solar cell structure based on mixed halide perovskite $CH_3NH_3PbI_{3-x}Br_x$ for values x=0,1,2, and 3 is simulated. As shown in Scheme 1, the cells' structure is designed according to the basic geometry of the thin film solar panels. In this model, a perovskite layer is sandwiched between two electron and hole transport layers (ETL,HTL) that are based on materials which have the capability to primarily transfer either electrons or holes due to a suitable positioning of the energy levels. Above them, a layer of crystalline indium tin oxide (ITO) is used as the anode with a thickness of 80nm because of its higher transmission coefficient, transparency (around 85%) and lower parasitic absorption. Also, at the bottom an Aluminum substrate is chosen as the cathode due to its high electrical conductivity. The selected materials in each layer and their thicknesses have been listed in order of top to bottom in Table 1. All the calculations are done by performing the finite difference time domain (FDTD) method provided by a common FDTD software package. In fact, by considering the simple geometry of the solar cell a two-dimensional optical simulation (X,Y) at room temperature (300 ˚k) is accomplished which in X-direction periodic boundary conditions and in Y-direction perfectly matched layer (PML) boundary conditions are employed. In addition, a plane wave source which is vertically polarized with a wavelength range of 300-1400nm is used to play the role of an AM1.5 direct (+circumsolar) spectrum. It should be noted that the optical constants (n,k) of ITO, $TiO_2$ and Al has given directly in the Edward-Palik solids handbook I – III [46] while the optical constants of polystyrene sulfonate (PEDOT:PSS) is just based on measurements made via spectroscopic ellipsometry on the Ref [5]. Also, for anti-reflection coating a single-layer of SiO$_2$ (n=1.45) is used on the surface that will increase the throughput of the system.

*Table 1: materials thickness for each layer.*

| Layer name | Material | Thickness (nm) |
|---|---|---|
| Glass cover | SiO$_2$ | 100 |
| Transparent conductive oxide | ITO | 80 |
| Hole transport layer | PEDOT:PSS | 15 |
| Active layer | Perovskite | 350 |
| Electron transport layer | TiO$_2$ | 15 |
| Metal contact | Aluminum | 120 |

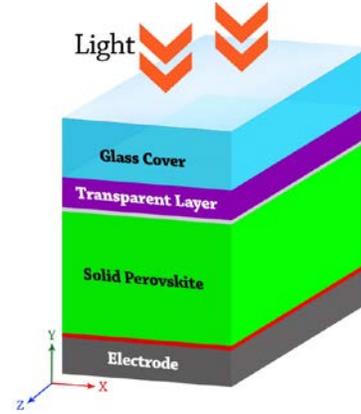

*Scheme 1: three-dimensional design of perovskite solar cell model.*

To determine the short-circuit current density (J$_{sc}$) and the light absorbance A(λ) of the system, FDTD solutions are utilized to compute the values of reflectance R(λ) and transmittance T(λ). Therefore, both quantity J$_{sc}$ and A(λ) can be calculated by following equations:

$$A(\lambda) = 1 - R(\lambda) - T(\lambda) \tag{3}$$

$$J_{sc} = \frac{e}{hc} \int \lambda\, A(\lambda)\, I_{AM1.5}(\lambda)\, d\lambda \tag{4}$$

where e is electron charge, h is the planck's constant, and c is the speed of light in vacuum while λ and I are the light wavelength and solar irradiance, respectively.

**DFT Results**

For investigation of structural properties of mentioned compounds, first band structures of pure perovskite (CH$_3$NH$_3$PbI$_3$) and doped compounds with bromide atoms (CH$_3$NH$_3$PbI$_{3-x}$Br$_x$) for amounts of x=0,1,2,3 are shown in symmetrical points of Brillouin zone in the Fig. 1. Band structures demonstrate that these structures have direct bandgap in cubic phase at the R point which with phase transition of perovskite and rotation of organic molecule bandgap shifts from R point to г point [17].

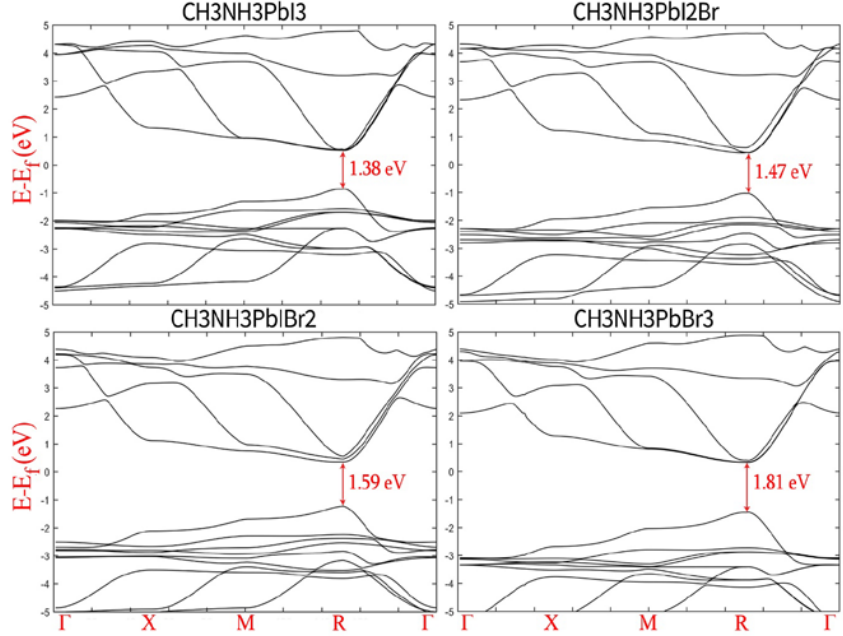

*Fig. 1. Band structure of perovskite compounds.*

Obtained results of band gap and lattice constants calculations express that increasing the amount of bromide in structure causes increase of the bandgap and decrease of the lattice constants. It comes from the fact that the ionic radius of bromide is smaller than iodine. Substitution of smaller atom in a bigger one causes increase of interaction between atoms in PbX6 octahedral and hybridization orbitals of bromide with lead is more than of iodine with lead [34]. Accordingly, we have to relocate the band energy in band structure to increase the bandgap. If the amount of doping raises, bandgap also increases which is shown in band structure of each compound. Our calculated band gaps are in a good agreement with previous theoretical works for MAPbI3 [7] and MAPbBr3 [15]. But DFT-GGA approximation underestimate the bandgap compared with experimental result [47,48]. The reason for this difference is excitonic effects that considered in experimental works while are absent in our calculation.

The reason of decreasing the lattice constants is doping of CH3NH3PbI3 by using bromide that is smaller than iodine halogen (Br<I) [34]. Also, in the investigations done on the doped compounds with chlorine($CH_3NH_3PbI_{3-x}Cl_x$), same results obtained, i.e. increasing of bandgap and reduction of lattice constants with increasing the chlorine in pure perovskite [30]. As it can be seen in Table 2, ionic radius of both Cl and Br is less than I, but Cl is smaller than Br, and has more difference with I. Therefore, changes in rate is greater in halide mixing compounds with Cl ($CH_3NH_3PbI_{3-x}Cl_x$). In other words, increasing of bandgap and reduction of lattice constants are more evident in $CH_3NH_3PbI_{3-x}Cl_x$ compounds (I > Br > Cl). In Scheme 2-unit cell of $CH_3NH_3PbI_3$ perovskite has been shown in cubic phase. Atoms 1, 2 and 3 are iodine halogens. First we put a Br instead of atom 1 ($CH_3NH_3PbI_2Br$), then another Br instead of atom2 ($CH_3NH_3PbIBr_2$) and eventually full halide $CH_3NH_3PbBr_3$. Table 3 is consisting of lattice constants and mean energy (total energy/number of atoms) of these perovskites. Substitution of Br in each direction causes lattice constants to decrease in the same direction. In $CH_3NH_3PbI_2Br$, bromide atom situates in the alignment of y-direction, $CH_3NH_3PbIBr_2$ in z-directions and also $CH_3NH_3PbBr_3$ situate in the alignment of x, y, z-direction. According to the mean energy values as shown in table 3, adding bromide to CH3NH3PbI3 perovskite lead to more structure stability and stronger connections.

*Table 2:* *Ionic radius of halogens (anions).*

| Halogen | I | Br | Cl |
|---|---|---|---|
| Ionic radius(Å) | 2.2 | 1.96 | 1.81 |

*Table 3:* *lattice constants and mean energy of considered perovskites.*

| Structure | a(Å) | b(Å) | c(Å) | E(eV/atom) |
|---|---|---|---|---|
| CH3NH3PbI3 | 6.31 | 6.31 | 6.31 | -129.91 |
| CH3NH3PbI2Br | 6.30 | 5.93 | 6.30 | -134.36 |
| CH3NH3PbIBr2 | 6.30 | 5.95 | 5.95 | -138.81 |
| CH3NH3PbBr3 | 5.95 | 5.95 | 5.95 | -143.27 |

In Fig. 2 density of states (DOS) of perovskite has been demonstrated. Increasing of bandgap with halide mixing is obvious. In fact, whenever embedment of bromide in compound increases the interaction and eventually hybridization of atom orbitals increases as well [34]. This makes the valance band edge to shift to lower energies and consequently increases the distance between maximum of valance band and minimum of conduction band.

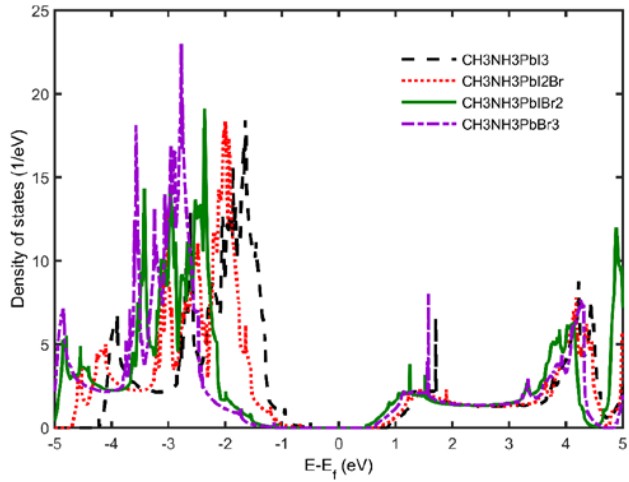
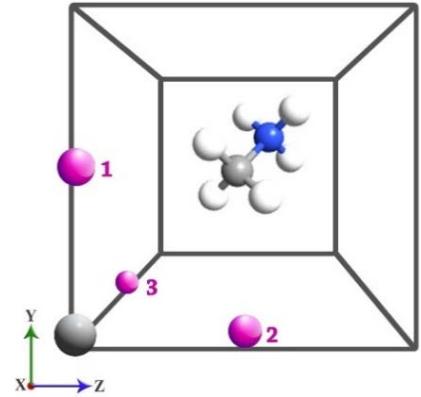

Scheme 2: unit cell of cubic $CH_3NH_3PbI_3$.

Fig. 2: Density of states of perovskite compounds.

For investigation of optical properties of these structures, their absorption coefficient, extinction coefficient and refractive index have been calculated and shown in Fig. 3(a, b, and c). Also for better understanding, the mixed halide compounds with chlorine have been considered too. In absorption coefficient diagram, full halide $CH_3NH_3PbBr_3$ and $CH_3NH_3PbCl_3$ do not show desirable absorption which the reason is their large bandgap. But the $CH_3NH_3PbBr_3$ has better absorption than $CH_3NH_3PbCl_3$. Other structures have nearly similar absorption, which demonstrate that they can be used instead of $CH_3NH_3PbI_3$ in designing of solar cells with more stability. Absorption coefficient has increased for energies higher than band gap. Low absorption is observed for energies under band gap which is due to computational error and negligible.

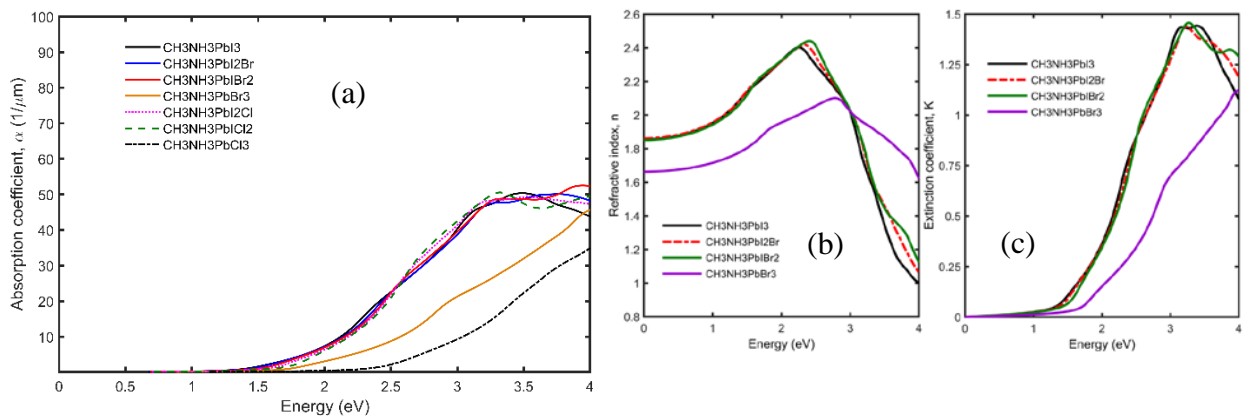

Fig. 3: a) Absorption coefficient b) Refractive index & c) Extinction coefficient of considered perovskites.

Our obtained results related to extinction coefficient and refractive index are consistent with theoretical and experimental investigations [49,50]. It is clear that optical properties of $CH_3NH_3PbBr_3$ is different from other three compounds. Other two doped compounds show properties similar to pure $CH_3NH_3PbI_3$. The same result is obtained for doped compounds with Cl [30]. The peaks seen in the extinction coefficient relate the inter-band transitions from maximum of valance band to minimum of conduction band. Bandgap absorption is due to transition of electron from first maximum valance band to first minimum conduction band at the R point. The peak located at the about 3, 3.5 eV, is related to inter-band transition at the X point from first maximum of valance band to first minimum of conduction band. But this peak in the band structure of $CH_3NH_3PbBr_3$ is due to transition at M point. Also the weak peak or the bump that is seen in $CH_3NH_3PbI_3$, $CH_3NH_3PbI_2Br$ and $CH_3NH_3PbIBr_2$ at about 2.5 eV energy, is referring to transition at the M point due to ignoring of excitonic contributions. However, it was appeared as a sharp peak in experimental works due to excitonic effects which are absent in our study [51,52].

**Investigation of the Solar Cell Optical Characteristics**

In this paper, we have considered mixed halide perovskite $CH_3NH_3PbI_{3-x}Br_x$ as a competitive substitute for $CH_3NH_3PbI_3$ since doping more bromine (Br) into the perovskite crystal will increase the stability of the structure of matter [31]. So at the beginning, we intend to evaluate the short-circuit current density of the solar cell by using these structures as an active area. The next step will be in an effort to improve this quantity. As can be seen in Table 4, the amounts of current density for AM1.5 Global spectra are given with this assumption the surface and bulk recombination of electrons and holes have been ignored. By a brief look at this table, we realize that the photon absorption process has been much better in $CH_3NH_3PbI_2Br$ and $CH_3NH_3PbIBr_2$ structures compared with the triiodide sample. It is quite evident that $CH_3NH_3PbIBr_2$ perovskite has the highest current density (39.2 mA/cm$^2$) which can be a preface to achieve high efficiency solar cells while no technique has been used for light trapping. In contrast, $CH_3NH_3PbBr_3$ compound has less chance for competition with the rest of structures. For the same cell, the ideal current density would be 30 mA/cm$^2$ when mixed halide perovskite $CH_3NH_3PbICl_2$ used as the active area. This difference is arising from more absorption of near-infrared light by $CH_3NH_3PbIBr_2$ perovskite. Since in our study $Br_2$ has had superior performance in comparison with other perovskites, from now on, all the investigations will be limited only for this structure.

*Table 4: The optical current density of cell for all perovskites.*

| Structure | $J_{sc}$ (mA.cm$^{-2}$) |
|---|---|
| CH3NH3PbI3 | 33.7 |
| CH3NH3PbI2Br | 37.8 |
| CH3NH3PbIBr2 | 39.2 |
| CH3NH3PbBr3 | 26.2 |

**Solar Cell Performance in the Presence of Plasmonic Array**

The light absorption rate of solar panels shows significant dependency on thickness of the active layer but thicker cells require greater amount of precious material. This obvious principle gives us the opportunity to enhance the light absorbance, by considering another method except light scattering structure or thicker absorbent layer. The noble metallic nanoparticles create surface plasmon through cumulative oscillations of free electrons [53]. These metallic nanostructures can increase significantly the optical path length of light in thin active layer and improve absorption of photons. In the meantime, silver nanoparticles have unique optical, electrical, and thermal properties and are being embedded in products that have a range from photovoltaics to chemical sensors. They are extraordinarily efficient at absorbing and scattering light and, unlike many dyes and pigments, have a color that depends upon the size and the shape of the particle [54]. In perovskite solar cells, the utilization of silver nanoparticles within the active region cause considerable increase in ideal current density and also light absorbance in the IR range [41]. In Scheme 3, the metallic arrays used inside the active area are able to help us in achieving our goal. As shown in this design, silver nanoparticles are employed in form of spherical, rectangular, and triangular so that all their parameters are optimized. The period (D) of all three arrays are equal and about 160nm while their height (H) are different. This value for rectangular arrays is nearly 60nm but for spherical and triangular arrays is considered to be 100nm. The reason why nanoparticles have been used within the active layer, is preventing the other layers from parasitical absorption. In these circumstances, the photons fortune to engage in perovskite layer will be higher than ITO film. It should be mentioned that in order to increase the accuracy of calculations, a mesh space as large as 0.002 micrometer has been considered under the plasmonic nanoparticles.

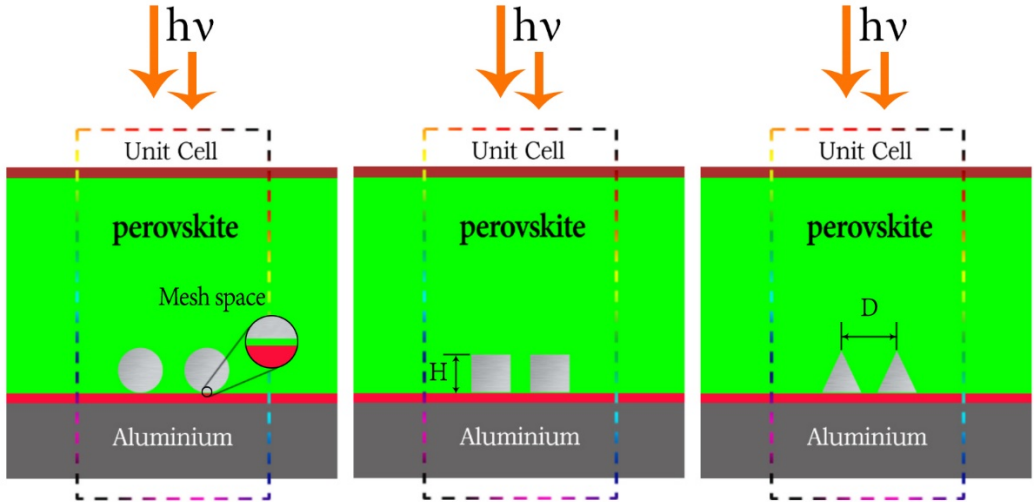

*Scheme 3: Side view of the thin film perovskite solar cell with three different nanoparticle arrays at rear surfaces; under normal incident light from the air.*

As the first result, in Fig. 4(a) the light reflectance curves are drawn in either presence or absence of the nanoparticle arrays. From this diagram can be found that the amount of light reflectance is virtually 24% without any nanoparticles, while has been reduced down to 13% with them. This acceptable status would happen only with the assistance of rectangular array, although triangular array can be handy too due to a reduction in reflectance rate down to 18%. By simply analyzing the results, the influence of particles geometry is obvious over total internal reflection. The arrays act as a light trapping structure which provide the terms of photons transfer and absorption by intensification of electromagnetic fields. In Fig. 4(b), the light absorption distribution inside the perovskite substrate have already been plotted accompanying their ideal current density (Jsc). These profiles indicate that at what levels the act of absorbing is more accomplished. From the diagram, it is obvious that the effective depth of absorption increases in the presence of silver arrangement. This effective depth is near to 235nm without plasmonic particles while the absorption is formed through overall thickness of perovskite (~350nm) with plasmonic particles. Actually, the lower energy light which is of a larger wavelength (such as infrared) is absorbed near the rear surface of solar cell. Furthermore, rectangular arrangement leads to an increment in cells current density from 39.2 to 43.7 mA/cm$^2$ that is remarkable. This design can be more practical for thicker cells.

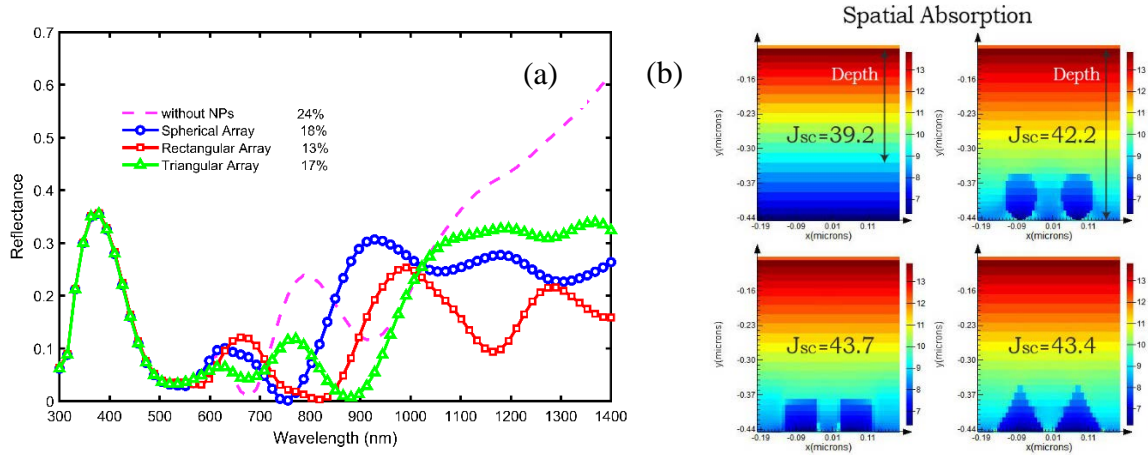

*Fig. 4: a) The light reflectance b) spatial absorption of perovskite solar cells in either presence or absence of silver nanoparticles arrays.*

Among other factors that comes with an impact on surface plasmon resonance, the genus of nanoparticles can be mentioned. Platinum (Pt), because of its unique features and extremely low absorption specifically at wavelengths larger than 300nm, is being used as a material for surface plasmon resonance effects. Pt nanoparticles show localized surface plasmon with higher sensitivity than silver nanoparticles [47]. So in this section, similar calculations were also accomplished on Pt nanoparticles just as the investigations on silvers. In Fig. 5(a), the light reflectance from top surface of the cell is plotted for Pt array. Similar to cells behavior with silver array in Fig. 4, reflected light is reduced considerably with a slight difference that in this case, Pt array causes trapping further light in the wavelength range between 700 to 1400nm. Actually, triangular array with almost identical performance compared with the rectangular array, cut way back the light reflectance to 6% which in turn will eventuate to a substantial enhancement in current density. Hence, the calculated current density attains to a high value of 47.9 mA/cm$^2$ from 39.2 mA/cm$^2$ (without any other NPs). Basically, it has experienced growth of 22%. The results indicate that Pt array own acceptable capability for light trapping in perovskite solar cells. Under these conditions, mixed halide perovskite $CH_3NH_3PbIBr_2$ can be called as a very good absorber in the visible and near-infrared region. Also, if you would like to see how Pt nanoparticles have influence on the amount of Jsc look at the Fig. 5(b-d). In this shape, the electric field lines are drawn within the unit cell that represent plasmonic particles have deflected the field lines. And that means the light beams have been scattered from their propagating direction. Obviously, the most deviation belongs to triangular form which prevents the light from penetrating to Al layer.

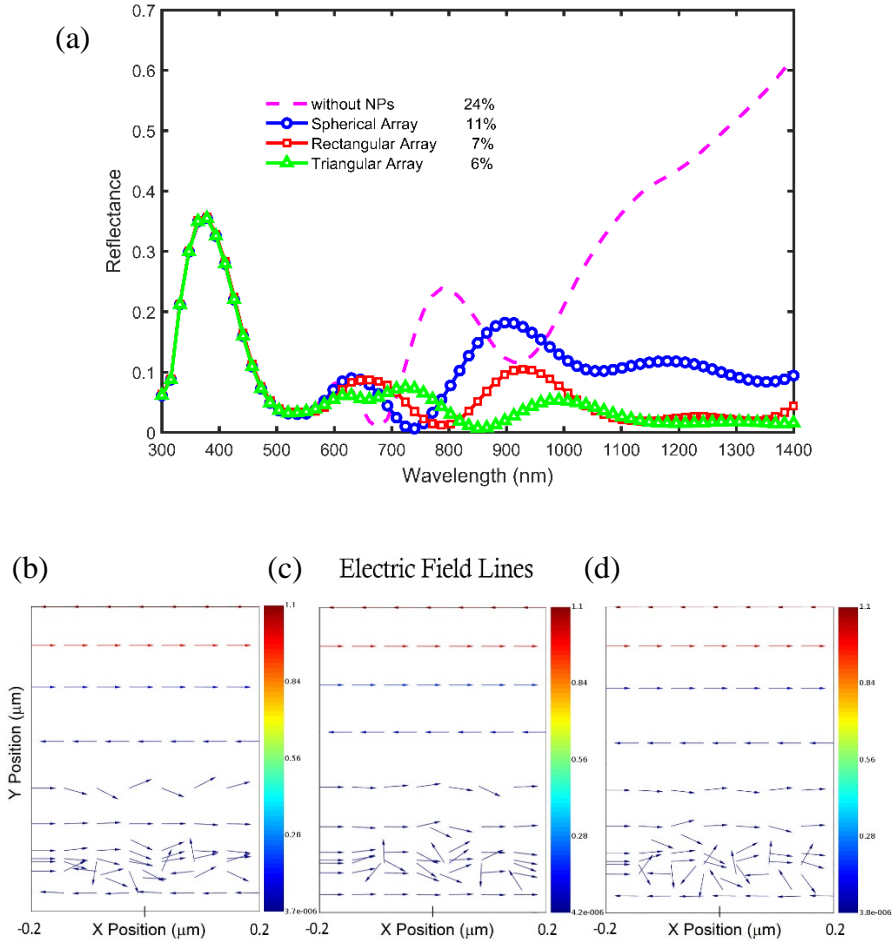

*Fig. 5: a) The light reflectance from top surface of the cell. Electric field lines inside the active layer in presence of b) spherical c) rectangular and d) triangular Platinum nanoparticles arrays.*

**Conclusions**

In this paper, initially the structural and optical properties of methylammonium lead iodide perovskite $CH_3NH_3PbI_3$ and doped compounds with halogen bromide $CH_3NH_3PbI_{3-x}Br_x$ in cubic phase have been analyzed. These calculations were carried out based on density functional theory (DFT) in order to observe the result of embedded halogen. Afterwards, a single-junction perovskite solar cell structure has been simulated at room temperature (300 °k). The results of DFT show that by increasing the proportion of Br in $CH_3NH_3PbI_3$ compound, the energy gap will be increased and lattice constants will be reduced. Also, from the results of simulation is seen that we can increase the ideal current density of solar cell as much as 22% by using a Platinum nanoparticles array which is very remarkable.